\documentclass[aps]{revtex4}
\usepackage{graphicx,epsfig}

\begin{document}
\title{Stability and hierarchy problems in string inspired braneworld scenarios}
\author{Saurya Das\footnote{E-mail: saurya.das@uleth.ca}$^a$,
Anindya Dey\footnote{E-mail: anindya@physics.utexas.edu}$^b$,
Soumitra SenGupta\footnote{E-mail: tpssg@iacs.res.in}$^c$}
\affiliation{$^a$Department of Physics\\
University of Lethbridge \\
4401 University Drive, Lethbridge\\
Alberta - T1K 3M4, Canada}
\affiliation{$^b$ The University of Texas at Austin \\
Theory Group, \\
1 University Station, C1608 \\
Austin, TX 78712-0269, USA}
\affiliation{$^c$Department of Theoretical Physics\\
Indian Association for the Cultivation of Science\\
Calcutta - 700032, India}
\begin{abstract}
We generalise the RS braneworld model by taking into account a
general stringy bulk containing the scalar dilaton field and the
two-form Kalb-Ramond field, apart from gravity.
Assuming small fluctuations around a RS background, the back-reacted
warp factor is obtained. It is shown that
the fine tuning problem in connection with the Higgs mass reappears in a
new guise and the effective modular potential
fails to stabilise the braneworld.\\
\end{abstract}

\maketitle

In recent years, braneworld models with extra spatial dimension(s) have
become popular as viable alternatives to supersymmetry as a means of
resolving the fine tuning problem (in connection with the large radiative correction to the
Higgs mass) in the Standard Model of elementary particles \cite{add,rs,arkani1}.
In the model proposed by Randall and Sundrum \cite{rs},
one considers
a $5$ dimensional anti de sitter spacetime with the extra spatial dimension
orbifolded as $S_1/Z_2$. Two $(3 + 1)$- dimensional branes, known as the
visible (TeV) brane and hidden (Planck)
brane are placed at the two orbifold fixed points, with the following
bulk metric ansatz:
\begin{equation}
ds^2=\exp{(-A)}\eta_{\mu\nu}dx^{\mu}dx^{\nu} - dy^2~,
\label{eq1}
\end{equation}
where $\mu,\nu=0,1,2,3$ (i.e. the visible coordinates) and $y=r\phi$ is the
extra spatial orbifolded coordinate. $r$ measures the distance between the two branes and
$\phi$ is the angular coordinate.
($\eta_{\mu\nu}$ is the usual $4$-dimensional Minkowski metric, whereas
$G_{MN}$ etc will denote the full five-dimensional metric).
In the original RS model, the Standard Model fields (open string excitations
attached to the brane) are assumed to be
localized on the
visible brane, whereas gravity (a closed string excitation)
propagates in the bulk. As a result of the warped background geometry,
all mass scales in the theory get exponentially warped
to the TeV scale, thereby resolving the hierarchy issue.
However, the stabilisation of this braneworld with a radius of compactification
of the order of Plank length has never been satisfactorily established.
The stabilising model using a bulk scalar field with interactions localised on
the branes \cite{gw} does not take into account the back-reaction of this bulk
field on the background geometry.
Other pieces of work which aimed at including
back-reaction \cite{backreaction} dealt with very special scenarios.\\

In a string-inspired scenario, it is well known that apart from gravity, the
two other massless closed string modes, namely scalar dilaton and the
two-form Kalb
Ramond (KR) field, can also propagate in the bulk \cite{burgess,ssg}.
Our present work aims to investigate both the fine tuning as well as the
modulus
stabilisation issues in presence of these bulk fields in a back reacted
$S_1/Z_2$ orbifolded geometry.
We determine an expression for the modified warp factor
with small fluctuations around the RS background. Such fluctuations
result from the backreaction of the bulk dilaton as well as the KR field,
within the perturbative regime.
It turns out that the hierarchy problem in connection with the scalar masses problem can be resolved
only if the background energy density of the KR field is fine tuned to an
unnaturally small value. This in a sense brings back the fine tuning problem
in a
new guise. Moreover we show that, despite the presence of a bulk scalar
in the form of dilaton, the braneworld modulus is intrinsically unstable.

In suitable units and in the Einstein frame, we begin with the RS metric ansatz (\ref{eq1}), and the action
\begin{eqnarray}
S &=& S_{Gravity} + S_{vis} + S_{hid} + S_{KR} + S_{dilaton}~, \\
\mbox{where,}~~~~S_{Gravity} &=& \int d^4x~d{y}~\sqrt{G}~[ 2M^3R + \Lambda]\\
S_{vis} &=&  \int d^4x \sqrt{-g_{vis}}~[L_{vis} - V_{vis}]\\
S_{hid} &=&  \int d^4x \sqrt{-g_{hid}}~[L_{hid} - V_{hid}]\\
S_{KR} &=& \int d^4x~d{y}\sqrt{G}~\exp({\Phi}/M^{3/2})~[ -2H_{MNL} H^{MNL}]\\
S_{dilaton}&=& \int d^4x~d{y}\sqrt{G}~\frac{1}{2}[\partial^{M}\Phi\partial_{M}\Phi - m^{2}~\Phi^{2}] ~.
\end{eqnarray}
Here $\Lambda$ is the five dimensional cosmological constant,
$V_{vis}, V_{hid}$ are the visible and hidden brane
tensions.
$H_{MNL} = \partial_{[M} B_{NL]}$
is the third rank antisymmetric field strength
corresponding to the two-form
KR field $B_{MN}$ \cite{ssg1}.
$\Phi$ is the scalar dilaton field present in the bulk.\\
Following \cite{gw} we also include interaction terms corresponding to the scalar dilaton
at the boundary.
\begin{equation}
S_{int}=-\int d^4 x \sqrt{-g_{SM}}\lambda_{s}(\Phi^2(y_{\pi})-v_{s}^2)^2
  -\int d^4 x \sqrt{-g_{Pl}}\lambda_{p}(\Phi^2(0)-v_{p}^2)^2
\label{int}
\end{equation} \\

The $5$ dimensional Einstein equations are as follows (where $'\equiv d/dy$):
\begin{eqnarray}
\frac{3}{2}{A'}^2&=&-\frac{\Lambda}{4M^3} -
\frac{1}{2M^3}
[3G^{\nu\beta}G^{\lambda\gamma}H_{y\nu\lambda}H_{y\beta\gamma}\exp(-\frac{\Phi}{M^{3/2}})
+ \frac{1}{4}(\Phi'^2 - m^2\Phi^2)]
\label{eq2} \\
\frac{3}{2}({A'}^2-{A''})&=&-\frac{\Lambda}{4M^3}-\frac{1}{2M^3}
[- 12\eta^{\lambda\gamma} H_{y 0 \lambda}H_{y 0 \gamma}+3G^{\nu\beta}G^{\lambda\gamma}
H_{y\nu\lambda}H_{y\beta\gamma}\eta{00}]\exp(-\frac{\Phi}{M^{3/2}})
+\frac{1}{8M^3}[\Phi'^2 + m^2 \Phi^2]
\label{eqna2}
\\
\frac{3}{2}({A'}^2-{A''})&=&-\frac{\Lambda}{4M^3}-\frac{1}{2M^3}
[- 12\eta^{\lambda\gamma} H_{y i \lambda}H_{y i \gamma}+3G^{\nu\beta}G^{\lambda\gamma}
H_{y\nu\lambda}H_{y\beta\gamma}\eta{ii}]\exp(-\frac{\Phi}{M^{3/2}})
+\frac{1}{8M^3}[\Phi'^2 + m^2 \Phi^2]
\label{eqna3}
\end{eqnarray}
In Eq.(\ref{eqna3}),
the index $i$ on the right hand side runs over 1,2 and 3, i.e. three spatial
components $x,y,z$, and there is no sum over $i$. Also, $\eta^{ij} \equiv g^{im} g^{jn} \eta_{mn}$.
Adding Eq.(\ref{eqna2}) and the $x,y,z$ components of Eq.(\ref{eqna3}),we get,
\begin{equation}
\frac{3}{2}[{A'}^2-{A''}]=-\frac{\Lambda}{4M^3} + \frac{1}{8M^3}[\Phi'^2 + m^2\Phi^2]
\label{eineqn6}
\end{equation}
\noindent
Subtracting Equ.(\ref{eineqn6}) from Equ.(\ref{eq2}),we have,
\begin{equation}
\frac{3}{2}A''=-\frac{1}{4M^3}\Phi'^2 -\frac{3}{2M^3}G^{\mu\nu}G^{\alpha\beta}H_{y\mu\alpha}H_{y\nu\beta}\exp{(-\Phi/M^{3/2})}\\
\label{eineqn7}
\end{equation}
The equation satisfied by the y-dependent VEV of the KR
fields is given as
\footnote{
It may appear that the solution for $H_{\mu\nu\lambda}$ in reference \cite{risi} (Eq.(7)) differs
from the one we have here. In that paper, $H_{\mu\nu\lambda}$ was expressed
as a dual of a vector field (Eq.(5)). Using this along
with Eq.(8), it can be shown however that the two solutions are consistent with each other.
}
,
\begin{equation}
G^{\mu\alpha}G^{\nu\beta}H_{y\mu\nu}H_{y\alpha\beta}=b M^5 \exp{(2A(y))}\exp{(2\Phi/M^{3/2})}
\label{eqm1}
\end{equation}
where $ b M^5 =\eta_{\mu\alpha}\eta_{\nu\beta}k^{\mu\nu}k^{\alpha\beta} $,
$k^{\mu\nu} $ is a constant antisymmetric tensor, independent
of $y$, and $b$ is  a dimensionless parameter measuring the energy density of the KR field.
It can be shown that the solution for $H_{\mu\nu\lambda}$ (or $B_{\mu\nu}$), derived from
the equation of motion for the KR field satisfies Eq.(14). The proof (without the dilaton field)
follows from our earlier paper \cite{ddsgproc}.
The proof including the dilaton field follows along similar lines.

Similarly, the classical equation of motion satisfied by the dilaton field
is given as,
\begin{equation}
\Phi''-2 A'\Phi'-m^2 \Phi^2 + \frac{6\exp{(-\Phi)}}{M^{\frac{3}{2}}}(G^{\mu\alpha}G^{\nu\beta}H_{y\mu\nu}H_{y\alpha\beta})=0
\label{eqm2}
\end{equation}
Now using Equ.(\ref{eqm1}) in Equ.(\ref{eqm2}),we have,
\begin{equation}
\Phi''-2 A'\Phi'-m^2 \Phi^2 + 6{M^{\frac{7}{2}}}b\exp{(2A)}\exp{(\Phi/M^{3/2})}=0
\label{eqm2a}
\end{equation}
Also using Equ.(\ref{eqm1}) in Equ.(\ref{eineqn7}), we have,
\begin{equation}
A''=-\frac{1}{6M^{3}}\Phi'^2
 -bM^2\exp{(2A)}\exp{(\Phi/M^{3/2})}
\label{eineqf}
\end{equation}
%
We linearise Eq.(16) and obtain the solution for
$\Phi$ as a power series in the dimensionless parameter $b$, which
is defined in Eq.(14). To leading order, the solution reads
\footnote{
In principle there could be non-perturbative solutions for which the
conclusions might be different. However in this paper, we examine the small
fluctuations produced by the KR and dilaton fields
(which may always be present), and how robust the resolution of the
hierarchy problem is under such fluctuations.}
:
\begin{eqnarray}
\Phi(\phi) &=& \Phi_0(\phi) + b\Phi_1(\phi)\\
&=&\Phi_0\exp{[2kr(1-\nu)\phi]} +
b\sum_{n=0}^\infty \frac{6M^{7/2}}{k^2}\frac{\left({\Phi_0/M^{3/2}}\right)^n}{n!}\frac{\exp{[kr\phi(2n(1-\nu)+4)]}}{(\omega_n^2 +
4\omega_n- m^2/k^2)}
\end{eqnarray}
%
Note that in order for this perturbation series to be valid over the
entire bulk spacetime, one requires
$b \lesssim \exp(-4kr\pi) \approx 10^{-64}$. In other words, the existence
of a perturbative solution around RS requires $b$ to be severely fine-tuned.
We now observe that in the above expression for $\Phi$, the leading order contribution from the summation in the RHS comes from
the term $n=0$. Substituting for $\nu$ with appropriate approximation, the truncated solution
(in the variable $y = r\phi$)
for $\Phi$ is obtained as,
\begin{equation}
\Phi(y) = \Phi_0 \exp[-m^2y/4k] - \frac{6bM^{7/2}}{m^2}\exp[4ky]
\end{equation}
Using the above solution for $\Phi$ we solve Equ.(\ref{eineqf}) for $A(y)$ to leading order in the perturbation
parameter $b$,
\begin{eqnarray}
A(y) &=& ky -\frac{\Phi_0}{M^{3/2}}\exp[-m^2y/2k] -32b\frac{\Phi_{0}M^{1/2}k^2}{(16k^2 -m^2)^2}\exp(4k-m^2/4k)y \nonumber \\
&-&bM^{2}\int\left[\int \exp\left[2ky + \frac{\Phi_0}{M^{3/2}}\exp[-m^2y/4k]\right]dy\right]dy
\label{backreact1}
\end{eqnarray}
Now since $\frac{\Phi_0}{M^{3/2}}\exp[-m^2y/4k] << 2ky $ in the exponent of the last term in the`
RHS, we can approximate the integrand
and obtain the following expression,
\begin{eqnarray}
A(y) &=& ky -\frac{\Phi_0}{M^{3/2}}\exp[-m^2y/2k] -32b\frac{\Phi_{0}M^{1/2}k^2}{(16k^2 -m^2)^2} \exp(4k-m^2/4k)y \nonumber \\
&-&\frac{bM^{2}\exp(\frac{\Phi_0}{M^{3/2}})}{(2k - m^2/4k)^2}\exp[(2k - m^2/4k)y]
\label{backreact2}
\end{eqnarray}
Equ.(\ref{backreact1}) or Equ(\ref{backreact2}) is the back-reacted expression for the warp factor $A(y)$ where the
second and the third terms in the
RHS are the contributions from the dilaton and the KR field respectively. It is easy to show that in absence of the dilaton
field we get back the expression for the warp factor in a KR-gravity bulk \cite{sdadssg}.\\
We now explore whether the KR-dilaton back-reacted warp factor can resolve
the hierarchy problem in connection with the mass of the Higgs boson. The scalar mass on the visible brane is given by
the warped relation,
\begin{equation}
m = m_0\exp[-A(y)]_{y=r\pi}
\end{equation}
where $m_0$ is the mass scale in the Planck brane.\\
We now estimate the contribution of mass warping due to the dilaton and KR field induced terms in the warp factor
(namely
the second, the third and the fourth terms in the RHS of Equ.(\ref{backreact2}) ).
Recall that in the original RS scenario $k$ and $r$ are
taken near the Planck mass and the Planck length to avoid the introduction of any unknown intermediate scale in the theory.
Our solution here [Equ.(\ref{backreact2})] describes a perturbative modification of the  warp factor over the RS value. Thus taking
$kr \sim 12$ with $k \sim M_{Pl}$ and $r \sim l_{Pl}$ along with $\Phi_0 \sim M^{3/2}$, we estimate the warp factor at
the visible brane as,
\begin{equation}
[A(y)]_{y =\pi} = 37 - 10^{-16} - b 10^{62} -b 10^{31}
\end{equation}
%
As our perturbative solution is valid for $b \lesssim 10^{-64}$,
the exponent $A(y)$ evaluated on the visible brane (including the
back-reaction) is always positive, and is very close to the RS value.
Thus, the above value of $b$, for which the perturbative expansion in
$\Phi$ is valid, results in a small fluctuation in the RS value of $A(y)$ in a
self-consistent manner. Therefore, the hierarchy problem can be resolved
even in the presence of the dilaton and KR fields, although the
parameter $b$ in the theory needs to be severely fine-tuned. We re-emphasise
that this fine-tuning arises from the requirement of the existence of
perturbative solutions to the equations of motion.
The desired warping from Planck to TeV scale therefore can be obtained only if the KR energy density parameter
$b$ is fine tuned to $10^{-60}$.
Thus the fine tuning problem reappears in a new guise. Similar fine tuning
was obtained in \cite{sdadssg} where only the KR field was considered in the bulk.\\
To understand the stability issue of this model we observe that
even without having to solve  the equations for $\Phi(y)$ and $A(y)$ explicitly,
it is possible to address the stabilisation issue with some very general assumptions
regarding the solution.
For the stabilisation analysis,we write down the complete form of the
stabilising potential, which is a function of $y_{\pi}$ (i.e the value of the extra coordinate at the
location of the visible brane),
\begin{eqnarray}
V_{\Phi}(y_{\pi})&=&\int_{0}^{y_{\pi}}dy\exp{-2A(y)}\exp{(-\Phi)}[-2H_{MNL}H^{MNL}]
+\exp{-2A(0)}\lambda_{p}(\Phi^2(0)-v_{p}^2)^2\nonumber \\
&+& \int_{0}^{y_{\pi}}dy\exp{-2A(y)}[\Phi'^2 + m^2 \Phi^2]
+ \exp{-2A(y_{\pi})}\lambda_{s}(\Phi^2(y_{\pi})-v_{s}^2)^2
\label{stab1}
\end{eqnarray}
Now,for the ground state configuration of $\Phi$
we take,
\begin{eqnarray}
\Phi(0)= v_{p}
\label{vac1}\\
\Phi(y_{\pi})=v_{s}
\label{vac2}
\end{eqnarray}
 Note that the solution $\Phi(\phi)$ contains only one constant $ \Phi_0$ .
Eliminating $\Phi_0$ from Equ.(\ref{vac1}) and Equ.(\ref{vac2}), we obtain
$y_{\pi}$ as a function of $v_{s}$ and $v_{p}$. Using this relation in
Equ.(\ref{vac2}), and using Equ.(\ref{vac1}) to eliminate $v_{p}$, we obtain
$\Phi_0$ purely as a function of $v_{s}$. This implies that the solution
$\Phi(\phi)$ has an explicit dependence only on $v_{s}$ and none on $y_{\pi}$.
Thus, the dependence of $V_{\Phi}(y_{\pi})$ on $y_{\pi}$ comes solely from the
upper limit of the integrals over the extra dimension.\\

In this form, the first and the second derivatives of the potential can be
readily obtained, as follows:
\begin{eqnarray}
V'_{\Phi}(y_{\pi})&=&\exp(-2A(y_{\pi})\exp{(-\Phi(y_{\pi})}[-2H_{MNL}H^{MNL}]_{y=y_{\pi}}
+ \frac{1}{2}\exp{(-2A(y_{\pi}))}[\Phi'^2 +
m^2 \Phi^2]_{y=y_{\pi}} \nonumber \\
&=&6\kappa \exp{(\Phi(y_{\pi}))} +\frac{1}{2}\exp{-2A(y_{\pi})}[\Phi'^2 + m^2 \Phi^2]_{y=y_{\pi}}
\label{stb2}
\end{eqnarray}
The second equality uses (\ref{eqm1}) with $\kappa= bM^5$.The condition of an
extremum requires $V'_{\Phi}(y_{\pi})=0$,the solution to which gives the
value of $y_{\pi}$ at which the braneworld is stabilised,viz.
\begin{equation}
 -6\kappa \exp{(\Phi(y_{\pi}))} =\frac{1}{2}\exp{(-2A(y_{\pi}))}[\Phi'^2 + m^2 \Phi^2]_{y=y_{\pi}}
\label{stb2a}
\end{equation}
The second derivative, on using the equations of motion and Equ.(\ref{stb2a}), gives,
\begin{equation}
V''_{\Phi}(y_{\pi})= \exp{(-2A(y_{\pi}))}\Phi'(y_{\pi})[4k\Phi'(y_{\pi}) + 2m^2 \Phi(y_{\pi})] + 12 \kappa \exp{\Phi(y_{\pi})}A'(y_{\pi})
\label{stb3}
\end{equation}
The above expression, whose sign determines the nature of the extremum, can
be calculated without resorting to the full solution of the equations of motion,
and using only the boundary values of $\Phi'$ and $A'$. Near the boundary at $y=y_{\pi}$
only the delta-function dependent terms in the equations of motion become important which could be
integrated to obtain the expressions of the first derivatives
at $y=y_{\pi}$. Thus one obtains,
\begin{eqnarray}
[\Phi'(y)]_{y_{\pi}} &=& -2\lambda_{s}\Phi(y_{\pi})(\Phi^2(y_{\pi})-v_{s}^2)=0
\label{bd1} \\
{[A'(y)]}_{y_{\pi}} &=& -\frac{1}{12M^3}V_{vis}
\label{bd2}
\end{eqnarray}
Using Equ.(\ref{bd1}) and Equ.(\ref{bd2}) in Equ.(\ref{stb3}),we find,
\begin{equation}
V''_{\Phi}(y_{\pi})= -\frac{\kappa}{M^3}\exp{\Phi(y_{\pi})}V_{vis}
\label{stb3a}
\end{equation}
As $V_{vis}$ is negative, the stability condition ( i.e $V'' \ge 0$) can be achieved
only if $ \kappa $ is positive. On the contrary, $\kappa$ is negative from
Equ.(\ref{stb2a}), if there exists a stationary point for the potential.
Thus, the presence of the stringy bulk fields back-reacts on the
geometry in a way which evidently jeopardises the stability of the braneworld.
However, it may be noted that if we consider the kinetic terms for the bulk dilaton
or KR field with an opposite sign (i.e a phantom like field ) then $\kappa$ will be
positive leading to a possible stability of the resulting braneworld.\\
To summarise,
we have shown that the requirement of fine tuning the Higgs mass by one part in $10^{16}$ can be avoided at the
expense of even more fine tuning of the KR field energy density
by one part in $10^{64}$.
Furthermore, inclusion of the dilaton and KR fields in the bulk results in an effective modular potential
which clearly does not have any minimum. Thus stabilisation of the modulus
cannot be achieved in presence of these stringy bulk fields. The modulus can however be stabilised
when phantom-like scalar fields are included in the bulk.
This work therefore raises questions about the efficacy of the string
inspired braneworld models to resolve the gauge hierarchy and the modulus stabilisation problems.

\vspace{.1cm}
\noindent
{\bf Acknowledgment}\\
The work of SD is supported by the Natural Sciences and Engineering Research Council of Canada.
SD and AD
acknowledges the hospitality of The Indian Association for the Cultivation of Science, Kolkata, India
for hospitality where a part of the work was done.
SSG acknowledges the hospitality of the University of Lethbridge, Canada,
during completion of this work.

\end{document}